\documentclass[APA,STIX1COL,numbers]{WileyNJD-v2}
\usepackage{moreverb}

\newcommand\BibTeX{{\rmfamily B\kern-.05em \textsc{i\kern-.025em b}\kern-.08em
T\kern-.1667em\lower.7ex\hbox{E}\kern-.125emX}}

\articletype{}%

\usepackage{macros_PCC_environmetrics} 
\usepackage{tikz}
\graphicspath{
  {../../../fig/social/simulation/}
  {../../../fig/social/killer_whales/}
}

\begin{document}

\title{Process convolution approaches for modeling interacting trajectories}

\author[1]{Henry R. Scharf}
\address[1]{Department of Statistics, Colorado State University
  Fort Collins, Colorado, USA}
\author[2,1]{Mevin B. Hooten}
\address[2]{U. S. Geological Survey,
  Colorado Cooperative Fish and Wildlife Research Unit,
  Department of Fish, Wildlife, and Conservation Biology, 
  Colorado State University,
  Fort Collins, Colorado, USA}
\author[3]{Devin S. Johnson}
\address[3]{Alaska Fisheries Science Center,
  National Marine Fisheries Service,
  National Oceanic and Atmospheric Administration, 
  Seattle, Washington USA}
\author[4]{John W. Durban}
\address[4]{Marine Mammal and Turtle Division, 
  Southwest Fisheries Science Center, 
  National Marine Fisheries Service, 
  National Oceanic and Atmospheric Administration, 
  La Jolla, California, USA}

\authormark{Henry R. Scharf \textsc{et al.}}




\corres{*Henry R. Scharf \\ 
  Department of Statistics \\
  Colorado State University \\
  Fort Collins, Colorado, USA. \\
\email{henry.scharf@colostate.edu}}


\abstract[Abstract]{ 
  Gaussian processes are a fundamental statistical tool used in a wide range of applications. In the spatio-temporal setting, several families of covariance functions exist to accommodate a wide variety of dependence structures arising in different applications. These parametric families can be restrictive and are insufficient in some situations. In contrast, process convolutions represent a flexible, interpretable approach to defining the covariance of a Gaussian process and have modest requirements to ensure validity. We introduce a generalization of the process convolution approach that employs multiple convolutions sequentially to form a ``process convolution chain.'' In our proposed multi-stage framework, complex dependencies that arise from a combination of different interacting mechanisms are decomposed into a series of interpretable kernel smoothers. We demonstrate an application of process convolution chains to model killer whale movement, in which the paths taken by multiple individuals are not independent, but reflect dynamic social interactions within the population. Our proposed model for dependent movement provides inference for the latent dynamic social structure in the study population. Additionally, by leveraging the positive dependence among individual paths, we achieve a reduction in uncertainty for the estimated locations of the killer whales, compared to a model that treats paths as independent.  
}


\jnlcitation{\cname{%
\author{Scharf H.}, 
\author{M. Hooten}, 
\author{D. Johnson}, and 
\author{J. Durban}, (\cyear{2017}), 
\ctitle{Process convolution approaches for modeling interacting trajectories}, \cjournal{Environmetrics}.}}

\maketitle


\section{Introduction}\label{sec:introduction}

Models for continuous random processes using kernel convolutions are known in much of the spatial and spatio-temporal statistical literature as ``process convolutions'' (e.g., \citealt{Higdon2002}; \citealt{Calder2007}; \citealt{Bolin2013}). Also referred to as spatial moving averages \citep[e.g.,][]{Cressie2002}, process convolutions first arose as a means for constructing valid covariance matrices for Gaussian processes (GPs), while relaxing the typical assumptions of stationarity and isotropy \citep[e.g.,][]{Barry1996}. One application area for process convolutions that has driven much of their recent theoretical development is for random spatial processes on stream networks \citep{VerHoef2006}. Stream networks occupy sub-manifolds of physical space on which specialized measures of distance often complicate enforcement of non-negative definiteness in covariance functions and have an inherent need for anisotropic dependence structures to accommodate the effects of direction in stream flow. Process convolutions offer an approach to modeling dependence that addresses both of these issues. We employ process convolutions to model the dependence observed in the trajectories of interacting animals.

A wealth of data arising from measurements of trajectories has spurred the development of many new statistical models for individual-based movement processes. These models have been used to study the individual movement patterns of a wide variety of animals, where scientific questions of interest are often focused on which exogenous environmental factors drive movement \citep{Potts2014}. A fundamental commonality of research related to animal movement is a need for accurate and precise knowledge about the paths traversed by each individual, leading to the development of many models used for reconstructing the true underlying movement processes from telemetry data \citep[e.g.,][]{Fleming2015}. Several excellent models for telemetry data have been proposed in both discrete- and continuous-time formulations that have sought to address the significant challenges making prediction of the true paths difficult (for a review see \citealt{McClintock2014}; \citealt{Hooten2016}). Telemetry devices are often subject to environmental conditions, and technological limitations that restrict their ability to deliver precise measurements of location at dense temporal resolutions. Thus, many analyses of animal movement must account for both measurement error and irregular temporal observations when estimating the true path traversed by an individual. In some cases, it may be reasonable to assume that the data include independent Gaussian noise; however, for many telemetry devices, it is necessary to account for complex, non-Gaussian forms of measurement error \citep{Brost2015, Buderman2016}. Irregularity in the frequency of telemetry observations can complicate the implementation of discrete-time models \citep[although see][for one solution using a multiple-imputation procedure]{Scharf2017} and potentially lead to large uncertainty during time intervals with few or no observations.

Ecologists are also interested in the extent to which individuals in a population move in direct response to each other. Particularly in the case of large mammals, complex and dynamic social connections within a population can play a critical role in the movement behavior of individuals (e.g., \citealt{Williams2006}; \citealt{Goldenberg2014}; \citealt{Scharf2016}). For example, two individuals with a strong social connection may exhibit similar movement patterns and visit the same locations, resulting in positive dependence between their respective paths. Inference about the social ties within a population, and how those ties change over time, can provide valuable information about the behavioral ecology of a population. However, there are few existing methods available that explicitly account for interactions among individuals (e.g., \citealt{Haydon2008}; \citealt{Codling2014}; \citealt{Langrock2014}; \citealt{Russell2016}; \citealt{Scharf2016}). Especially when direct observation of a species is infeasible, careful analysis of data gathered from telemetry devices can reveal useful information about a population's social structure \citep{Scharf2016}.

We propose a joint model for the movement of multiple individuals that expands the suite of statistical methodology available for studying animal movement. Our approach allows researchers to study dependence among the paths of a population of individuals that arises as the result of social interactions. Also, by taking into account the dependence among individuals in a population, our approach has the potential to reduce both bias and uncertainty in reconstructing trajectories in the study population compared to models that treat individuals independently. We obtain improvements in path reconstruction by modeling the true movement processes of the population conditioned on an unobserved, dynamic social network. The inferred network provides a description of the social ties within a population and how those ties change in time.

In what follows, we present a generalized approach to process convolutions in which a random process is decomposed into a sequence of one or more smoothing kernels that are convolved with a white noise process. Multiple stages of smoothing allow us to model dependence in time and between individuals separately, resulting in a multivariate Gaussian process that captures the combination of effects. It is important to note that, while the kernels responsible for inducing temporal and path-wise dependence are constructed separately, the resulting covariance function is not separable in the geostatistical sense.

We provide a detailed introduction to our generalized approach to process convolutions in Sections \ref{sec:PCC} and \ref{sec:reduced_rank} and construct our joint model for movement in Sections \ref{sec:dependent_movement} and \ref{sec:network_prior}. After discussing the implementation details in Section \ref{sec:implementation}, we demonstrate path reconstruction in a simulation study in Section \ref{sec:simulation_study}. We apply our model to the study of telemetry data arising from the movement of killer whales in Section \ref{sec:killer_whales} and close with a summary of our findings and future directions.

\section{Methods}\label{sec:methods}

\subsection{A multiple-kernel convolution framework}\label{sec:PCC}

We first outline a new flexible framework for the development of Gaussian process models based on a synthesis of ideas from geostatistics, multivariate time series, and trajectory modeling. Our hierarchical framework relies on the kernel convolution approach for modeling random processes, known, in the spatial and spatio-temporal statistical literature, as ``process convolutions'' (e.g., \citealt{Higdon2002}; \citealt{Calder2007}; \citealt{Bolin2013}). A mean zero random process $\mu(\cdot)$ is called a process convolution if it is constructed by convolving a continuous random process, $d\! B(\cdot)$, with a kernel function, $h$, over a domain, $\cT$, so that
\begin{align}
  \mu(t) &= \int_\cT h(t, \tau) d\! B(\tau). \label{eqn:pc}
\end{align}
If the random process $d\! B(\cdot)$ is Gaussian, the resulting process $\mu(\cdot)$ will also be Gaussian with covariance function 
\begin{align}
\Cov\lp \mu(t), \mu(t^*) \rp &= \int_\cT \int_\cT h(t, \tau) 
  h(t^*, \tau^*)
  \Cov \lp d\! B(\tau), d\! B(\tau^*) \rp. \label{eqn:cov_gen} 
\end{align}
It is common to define the process $d\! B(\cdot)$ to be Gaussian white noise, which yields the simplified covariance function
\begin{align}
  \Cov\lp \mu(t), \mu(t^*) \rp &= \int_\cT h(t, \tau) h(t^*, \tau) d\tau \label{eqn:cov}
\end{align}
(although see \citet{Nychka2015} for an example of a spatial process convolution where $d\! B(\cdot)$ has covariance specified through a Gaussian Markov random field). A process convolution with kernel function $h(\cdot, \cdot)$, as in \eqref{eqn:pc}, represents a smoothing of the process $d\! B(\cdot)$, and the kernel is therefore often referred to as a smoother. In what follows, we refer to $h$ as both a kernel function and a smoother, interchangeably.

The function defined in \eqref{eqn:cov} is guaranteed to be non-negative definite, and therefore a valid covariance function, if $\int_\cT h(t, \tau) d\tau < \infty$ and $\int_\cT h^2(t, \tau) d\tau < \infty$ for all $t$ \citep{Higdon2002}. In many applications, it may be easier to specify the proper form of dependence in a GP through the form of a kernel smoother rather than that of a covariance function. For example, by constructing process convolutions with asymmetrical kernels, \citet{VerHoef2006} accounted for directional dependence in spatial processes arising on stream networks, where directional flow plays a crucial role. For certain choices of $h$, it is possible to express the covariance function analytically (e.g., \citealt{Higdon2002}; \citealt{Paciorek2003}; \citealt{VerHoef2010}), although this is not necessary for non-negative definiteness. In the cases where an analytic solution to \eqref{eqn:cov} is not available, one can evaluate the integral numerically. The class of covariance functions constructed using kernel convolutions is general, containing many of the parametric families commonly used in geostatistical settings (e.g., exponential, Gaussian, spherical), as well as more flexible dependence structures. For example, \citet{Higdon1998} and \citet{Paciorek2003} showed that process convolutions allow researchers to model both anisotropic and non-stationary dependence by letting $h(t, \tau)$ vary with both $|t - \tau|$ and $t$. In principle, any covariance function evaluated over a discretized domain has a process convolution representation on the same discretized grid, where the kernel function's values over the grid may be defined using a decomposition of the covariance matrix (e.g., Cholesky). 

\citet{Hooten2016} used process convolutions to develop new models for trajectories by convolving a Wiener process, rather than white noise, with different kernel functions. Using Wiener process convolutions, they specified realistic models for animal movement without the characteristic ``roughness'' found in Brownian motion and define a framework that incorporated other existing models for movement, such as that in \cite{Johnson2008}. Brownian motion itself can be thought of as a process convolution where the kernel is a step function given by $h^{(bm)}(t, \tau) \equiv \bone{\tau \leq t}$ and the process is Gaussian white noise. Thus, the \citet{Hooten2016} framework can be viewed as a nested smoothing procedure, where the position at time $t$ is
\begin{align}
  \mu(t) &= \int_\cT h(t, \tau) \lp \int_\cT h^{(bm)}(\tau, \tiltau) d\! B(\tiltau) \rp d\tau. \label{eqn:hooten}
\end{align}
 To account for Brownian motion with an initial position far from the origin, one may define the process $d\! B(\cdot)$ to be a continuous white noise process for all $t>0$, with arbitrary initial variance so that $d\! B(0) \sim \N \lp 0, \sigma_0^2 \rp$. \citet{Hooten2016} provided several examples of possible kernel functions and demonstrated a computationally efficient procedure to incorporate temporal non-stationarity into the kernel, allowing them to fit highly flexible models to telemetry data arising from animal movement trajectories.

It is possible to write the two-stage process convolution in \eqref{eqn:hooten} in terms of a single effective smoothing kernel $\tilde{h}(t, \tiltau) = \int_\cT h(t, \tau)h^{(bm)}(\tau, \tiltau) d\tau$ using Fubini's theorem to change the order of integration. Moreover, the equivalent, single-stage representation for $\mu(\cdot)$ is not limited to two-stage smoothing processes. One may specify any number of kernel functions and the resulting multi-stage kernel convolution can be written in terms of a single effective kernel. This nested structure motivates a novel model-building approach in which an arbitrarily large collection of convolution kernels is used to specify a valid GP.  We refer to a stochastic process resulting from the iterative convolution of a chain of kernels $h^{(1)}, \dots, h^{(L)}$ as a ``process convolution chain'' (PCC) and write it in its expanded form as
\begin{align}
  \begin{split}
    \mu^{(1)}(\tau_1) &= \int_\cT h^{(1)}(\tau_1, \tau_0) d\! B(\tau_0) \\
    \mu^{(\uline{2})}(\tau_2) &= \int_\cT h^{(2)}(\tau_2, \tau_1) d\mu^{(1)}(\tau_1) \\
    \vdots \\
    \mu^{(\uline{L})}(\tau_l) &= \int_\cT h^{(L)}(\tau_L, \tau_{L-1}) d\mu^{(\uline{L-1})}(\tau_{L-1}).
  \end{split} 
  \label{eqn:pcc}
\end{align}
The underline in stages $2 \leq l \leq L$ (i.e., $\mu^{(\uline{l})}(\tau_l)$) distinguishes the PCC composed of all kernels $\lbr h^{(k)}: k \leq l \rbr$ from that of convolving $h^{(l)}$ with the process $d\! B(\cdot)$ directly (written $\mu^{(l)}(\tau_l)$). We use matching superscripts to denote the covariance function for a given process (e.g., $\Cov(\mu^{(\uline{l})}(t),\mu^{(\uline{l})}(t^*)) = C^{(\uline{l})}(t, t^*)$). A collapsed version of a PCC using a single effective kernel and the white noise process $d\! B(\cdot)$ can be written as
\begin{align}
  \mu^{(\uline{l})}(\tau_l) &= \int_\cT h^{(\uline{l})}(\tau_l, \tau_0) d\! B(\tau_0) \label{eqn:convolution_chain} 
  \intertext{if we define the effective kernel as} 
  h^{(\uline{l})}(\tau_l, \tau_0) &\equiv \int_\cT \dots \int_\cT h^{(l)}(\tau_l, \tau_{l-1}) \dots
  h^{(1)}(\tau_1, \tau_0) d\tau_1 \dots d\tau_{l-1}.
\end{align}
To simplify notation, we write $\tilde{h} = h^{(\uline{L})}$ for the effective kernel constructed from the entire chain, $\tilmu$ for the full process $\mu^{(\uline{L})}$, and $\tilC$ for the corresponding covariance function. In general, the kernels $h^{(k)}$ are not commutative, in the sense that
\begin{align}
  \int_\cT \int_\cT h^{(l_1)}(t, \tau) h^{(l_2)}(\tau, \tau_0) d\! B(\tau_0) \neq \int_\cT \int_\cT h^{(l_2)}(t, \tau) h^{(l_1)}(\tau, \tau_0) d\! B(\tau_0).
\end{align}
Thus, both the forms of the kernels and their order are important when specifying a PCC.

Specifying a GP through an ordered chain of smoothers allows for considerable flexibility. For example, if we specify a chain of size $L = 1$ with kernel function $h^{(1)}(t, \tau_1) \equiv \bone{\tau_1 < t}$, we recover simple Brownian motion. If we increase the length of the chain by including a second kernel, $h^{(2)}(t, \tau_2) = (1 + \frac{\tau_2 - t}{\phi})\bone{-\phi < \tau_2 - t \leq 0}$, we recover a structure used by \citet{VerHoef2010} for modeling dependence in stream networks \citep[also mentioned in][Figure 2, row 3]{Hooten2016}. Another specific case of PCCs are so-called $(k-1)$-fold integrated Wiener processes \citep{Shepp1966, Wecker1983, Rue2005a}, which use $L = k-1$ kernels of the form $h^{(l)}(t, \tau) = (t - \tau)/(l - 1)$ and allow one to model GPs with exactly $m$ continuous derivatives. The multi-stage decomposition is similar in spirit to the way random processes with multiple dependence scales are decomposed additively in multi-resolution processes (\citealt{Higdon2002}; \citealt{Nychka2015}; \citealt{Katzfuss2016}); the difference is the convolution chain approach combines model components using convolution rather than addition.

In what follows, we construct convolution kernels that allow us to specify multivariate GPs; these provide a powerful tool for modeling multiple trajectories arising from interacting individuals. We return to this specific application of PCCs in Section \ref{sec:dependent_movement}.

\subsection{Finite representation}\label{sec:reduced_rank}

The joint distribution of $\tilbmu \equiv \lp \tilmu(t_1), \dots, \tilmu(t_n) \rp'$ for $\bt \equiv \lp t_1, \dots, t_n \rp' \subseteq \cT$ is mean-zero Gaussian with covariance function $\tilC(t, t^*) = \int_\cT \tilde{h}(t, \tau) \tilde{h}(t^*, \tau) d\tau$ for any $t, t^* \in \bt$ (if the integrals of $\tilh$ and $\tilh^2$ exist). When this integral cannot be computed analytically, one can perform the integration numerically by selecting a fine grid of values $\btau \equiv \lp \tau_0, \dots, \tau_m \rp'$ from the domain $\cT$, and evaluating the discrete sum $\sum_{i=1}^m \tilde{h}(t, \tau_i)\tilde{h}(t^*, \tau_i) \Delta\tau_i$, where $\Delta\tau_i = \tau_i - \tau_{i - 1}$. In the discretized setting, the process convolution may be written as a matrix product $\tilbmu = \tilbH \bep_{0}$, where the $j^{th}$ row of $\tilbH_{n \times m}$ is the function $h(t_j, \tau)$ evaluated at all $\tau \in \btau$, and $\bep_{0} \equiv \lp \varepsilon_{0}(\tau_1), \dots, \varepsilon_{0}(\tau_m)\rp'$ with each $\varepsilon_{0}(\tau_i) \sim \N(0, \Delta\tau_i)$. Assuming the grid times $\btau$ are equally spaced with intervals of size $\Delta\tau$, the joint distribution of $\tilbmu$ can be expressed using matrix notation as
\begin{align}
  \tilbmu \sim \N \lp \bzero, \Delta \tau \tilbH \tilbH' \rp. \label{eqn:discrete_second_order}
\end{align}
As the density of the grid grows to infinity ($m \rightarrow \infty$), the outer product in \eqref{eqn:discrete_second_order} approaches the covariance defined by the continuous integral over $\cT$. The granularity of the grid (i.e., the size of $m$) required to adequately approximate the integral will generally depend on the characteristics of $\tilde{h}$. Alternatively, one can follow the approach of \citet{Higdon2002}, who used finite process convolutions \citep[also called discrete process convolutions in][]{Calder2008} as an approximation, and choose $m < n$ to yield a fixed-rank model for the continuous process. Fixed-rank models can offer considerable computational efficiency and provide an adjustable level of implicit regularization \citep{Wikle2010}.

\subsection{A process convolution chain for dependent movement}\label{sec:dependent_movement}

\subsubsection{Social smoothing}\label{sec:social_smoothing}

We construct a novel model for animal movement using the PCC approach within a Bayesian hierarchical modeling framework. To describe the basic procedure, we first consider paths in one dimension. The model we describe can be readily extended to movement in two or more dimensions, and we demonstrate that in the example that follows. We use a three-stage ($L = 3$) PCC that includes a specialized kernel function constructed to induce dependence among the paths of $p$ different individuals. The inter-path dependence that arises is based on a weighted, undirected, latent social network \citep{Goldenberg2010}.

Let the random variable $\mu_i(t)$ represent the position of individual $i$ at time $t$, and let $w_{ij}(\tau) \in \lb 0, 1 \rb$ denote the connection weight between individuals $i$ and $j$ at time $\tau$, with $w_{ii}(\tau) \equiv 1$ and $i, j \in \lbr 1, \dots, p \rbr$. We propose a ``social'' smoothing kernel of the form
\begin{align}
  h^{(soc)}_{ij}(\tau_{soc}, \tau) &\equiv \bone{\tau = \tau_{soc}}
  \frac{w_{ij}(\tau_{soc})}{| w_{i\cdot}(\tau_{soc}) |}, \label{eqn:soc_kernel} \\
  | w_{i\cdot}(\tau_{soc}) | &\equiv \sum_{j=1}^p w_{ij}(\tau_{soc}).
\end{align}
We write the kernel in \eqref{eqn:soc_kernel} with explicit dependence on arbitrary time $\tau$ for completeness, but will hereafter suppress the second argument for brevity. To provide intuition for the effect of convolving with the social smoothing kernel, we first consider the case of a multivariate process with two individuals. We assume that, for the process $\bmu(\cdot) \equiv \lp \mu_1(\cdot), \mu_2(\cdot) \rp'$, the variables $\mu_1(\tau)$ and $\mu_2(\tau^*)$ are independent $\textit{a priori}$ for all $\tau$ and $\tau^*$. After social smoothing, the resulting process for individual 1 is defined by
\begin{align}
  \mu_1^{(soc)}(\tau_{soc}) &\equiv \sum_{j = 1}^2
  \int_\cT h^{(soc)}_{1j}(\tau_{soc}, \tau) \mu_j(\tau) d\tau \\
  &= h_{11}(\tau_{soc})\mu_1(\tau_{soc}) + h_{12}(\tau_{soc})\mu_{2}(\tau_{soc}) \label{eqn:social_smoothing1}\\
  &= \frac{\mu_1(\tau_{soc})}{1 + |w_{12}(\tau_{soc})|} + 
  \frac{w_{12}(\tau_{soc})\mu_{2}(\tau_{soc})}{1 + |w_{12}(\tau_{soc})|}. \label{eqn:social_smoothing2}
\end{align}
From \eqref{eqn:social_smoothing1} and \eqref{eqn:social_smoothing2}, it is clear that $\mu_1^{(soc)}(\tau_{soc})$ is a weighted average of the independent variables $\mu_1(\tau_{soc})$ and $\mu_2(\tau_{soc})$, where the weights are equal to $h_{11}(\tau_{soc})$ and $h_{12}(\tau_{soc})$, respectively. When the network connection is strong (i.e., $w_{12}(\tau_{soc}) \approx 1$), the two weights $h_{11}(\tau_{soc})$ and $h_{12}(\tau_{soc})$ will be approximately equal, and the effect of the kernel is to ``squeeze'' the previously independent paths toward a scaled version of their mutual mean. When the network connection is weak ($w_{12}(\tau_{soc}) \approx 0$), the weights will be close to 1 and 0 respectively, and the processes will be unaffected by the kernel, retaining the \textit{a priori} independence between individuals. These effects are visible in the simulated process shown in the top plot of Figure~\ref{fig:SPCC}. 

The social smoothing kernel we propose is motivated by an interest in describing a mechanistic, interpretable driver of dependent movement. Specifically, it is useful for modeling the paths of individuals who have a tendency to move toward, and alongside, other individuals with whom they share connections. This type of dependence is ubiquitous in animal movement, but is by no means the only meaningful type of interaction. As an alternative, consider the movement of two highly territorial animals with overlapping territories in which observed paths would appear to ``avoid'' each other. To model mutual avoidance, we require a social kernel that ``spreads apart'' two otherwise independent paths when a strong social connection is present. In what follows, we focus on the particular kernel defined in \eqref{eqn:soc}, but we emphasize that PCCs, with suitably constructed kernels, offer a flexible way to specify models for a wide range of possible behavioral mechanisms. Additionally, our choice of social smoothing kernel is most appropriate when social connections among individuals vary slowly relative to the movement processes. When relationships among individuals are allowed to vary too rapidly, it may become difficult for the model to distinguish between brief encouters that arise due to social effects, and those that arise as part of the stochasticity inherent in Brownian motion.

Equipped with the social kernel given in \eqref{eqn:soc}, we outline the first two stages of our full three-stage PCC model for the dependent movement of $p$ individuals. We begin with Gaussian white noise and smooth at the first stage convolving $h_{ij}^{(bm)}(t, \tau) \equiv \bone{\tau \leq t}\bone{i=j}$ with $d\! B(\cdot)$ to generate $p$ independent instances of Brownian motion, denoted $\mu_i^{(bm)}(\cdot)$, each with its own initial position. At the second stage, we apply the social smoothing kernel defined in \eqref{eqn:soc_kernel} to the collection of all $p$ processes, $\bmu^{(bm)}(t) \equiv \lp \mu_1(t), \dots, \mu_p(t) \rp'$. Smoothing with the social kernel returns weighted averages of the Brownian processes, where the weights are proportional to the latent social weights $w_{ij}(t)$. Thus, two individuals $i$ and $j$ for whom $w_{ij}(t)$ is close to 1 will tend to have smoothed locations $\mu^{(\uline{soc})}_i(t)$ and $\mu^{(\uline{soc})}_j(t)$ that are close together in space. A third and final stage in the PCC, introduced in the following section, ensures that the random process has the proper temporal smoothness required for modeling animal movement.

\subsubsection{Inertial smoothing} \label{sec:inertial_smoothing}

Marginally, the individual paths $\bmu_i^{\uline{soc}}$ generated from the two-stage PCC constructed from $h^{(bm)}$ and $h^{(soc)}$ are each an instance of Brownian motion. However, Brownain motion is unsuitable for direct modeling of animal movement because the instantaneous velocity of a particle traversing a Brownian path is discontinuous, and therefore the acceleration is not well defined. Discontinuities in the first derivative of the path processes imply that individuals are capable of instantaneous changes in their velocities and is inconsistent with the physical laws governing the mechanics of massive bodies \citep[e.g.,][]{Feynman1963}.

Several process convolutions have been explored that impart specific smoothness properties on a GP. For example, \citet{Shepp1966}, \citet{Wecker1983}, and \citet{Rue2005a} discuss applications of $(k-1)$-fold integrated Wiener processes, which ensure the existence of $k - 1$ continuous derivatives. \citet{Johnson2008} constructed a model for the movement of individual harbor and northern fur seals by specifying an Ornstein-Uhlenbeck process for the velocity, rather than the position. \citet{Hooten2016} modeled the true locations of an individual animal using a two-stage PCC in which the second kernel function is Gaussian, and \cite{Buderman2016} used a similar model with cubic splines as their kernels. In all of these examples, smoothness is imparted to a process through convolution with a kernel, either implicitly or explicitly. We introduce a final stage of smoothing to yield paths that are guaranteed to be continuously differentiable.

\begin{figure}[htbp]
  \centering
  \includegraphics[width = \textwidth]{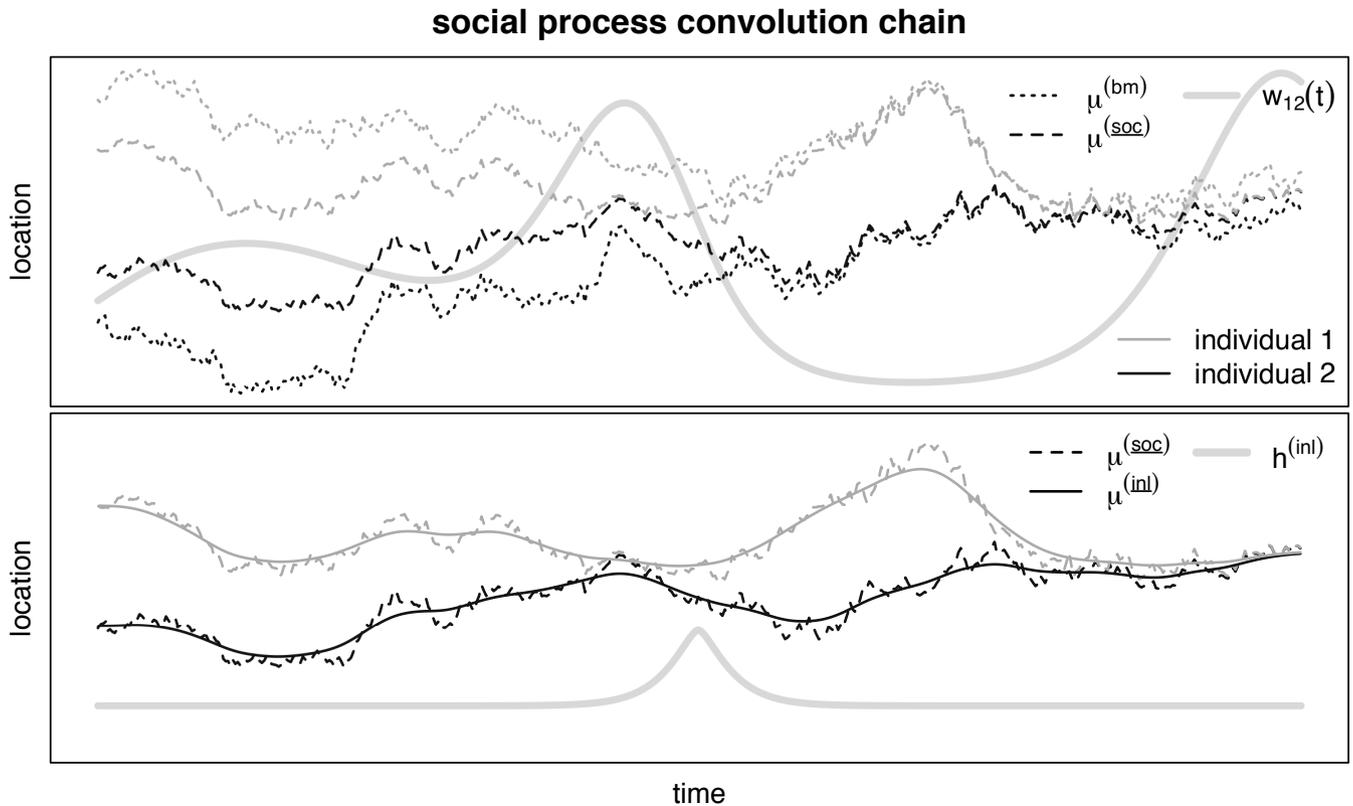}
  \caption{An example of a one dimensional GP arising from a three-stage social PCC for two individuals (light and dark). All three intermediate stages of smoothing are shown, beginning with Brownian motion (top, dotted), followed by ``social'' smoothing (top and bottom, dashed), and finally ``inertial'' smoothing (bottom, solid). The thick gray line in the top plot is proportional to the strength of the social tie $w_{12}(t)$ at all times $t$, where the extent of the y-axis corresponds to a range of $\lb 0, 1 \rb$. The thick gray line in the bottom plot is proportional to the inertial smoothing kernel $h^{(inl)}(t)$ for a fixed value of $\tau$.
  \label{fig:SPCC}}
\end{figure}

Enforcing the existence of a continuous derivative for the movement processes ensures that individuals have an interpretable acceleration at every time $t$. Therefore, we refer to the final kernel function as an ``inertial'' smoother because its purpose is to generate paths with the physical properties required to obey the laws of classical mechanics. We specify a kernel from the Mat\'ern family of correlation functions \citep{Cressie1991} with an unknown range parameter, $\phi_{inl}$, and smoothness $\nu = 1$ as the inertial smoother. The effect of this final kernel, $h^{(inl)}_{ij}(t, \tau) \equiv \frac{|\tau_{inl} - \tau_{soc}|}{\phi_{inl}} K_1\lp |\tau_{inl} - \tau_{soc}| / \phi_{inl} \rp \bone{i=j}$, where $K_1$ is a modified Bessel function, is visible in the bottom plot of Figure~\ref{fig:SPCC}. As we did for the social smoothing kernel, in what follows we suppress the superfluous index and write $h^{(inl)}(\cdot)$.

\subsubsection{The full three-stage process convolution chain} \label{sec:three_stage_PCC}

The expanded specification of the full three-stage PCC is given by
\begin{align}
  \mu^{(bm)}_i(\tau_{bm}) &= \int_\cT h^{(bm)}(\tau_{bm}, \tau_0)
  d\! B_i(\tau_0) \label{eqn:bm} \\
  \mu_i^{(\uline{soc})}(\tau_{soc}) &= \sum_{j = 1}^p h^{(soc)}_{ij}(\tau_{soc})
  \mu^{(bm)}_j(\tau_{soc}) \label{eqn:soc} \\
  \tilmu_i(\tau_{inl}) = \mu^{(\uline{inl})}_i(\tau_{inl}) &= \int_\cT h^{(inl)}(\tau_{inl}, \tau_{soc})
  \mu_i^{(\uline{soc})}(\tau_{soc}) d\tau_{soc}. \label{eqn:smooth}
\end{align}
One of the primary benefits of using a PCC framework is that it allows one to decompose complex dependencies into an iterative sequence of relatively simple mechanisms. Positions of multiple individuals can, in general, exhibit dependencies in both time and among individuals, and the characteristics of these dependencies may be dynamic themselves. We construct a PCC using kernels that compartmentalize the mechanisms generating the complex dependencies in the collective movement of interacting individuals. The first stage of smoothing is in time, and not across individuals, so that $\mu_i^{(bm)}(\tau_{bm})$ and $\mu_j^{(bm)}(\tau_{bm})$ are independent for $i\neq j$. The second stage of smoothing is across individuals, and not in time, in the sense that $h^{(soc)}_{ij}(t, t^*) = 0$ for $t \neq t^*$. The final stage of smoothing is in time only.

Restricting individual smoothing components to operate in only one ``dimension'' at a time (in this case, either temporal or social) is not necessary. However, a primary motivation for constructing GP models in the PCC framework is to decompose a complex mechanism into components that are easier to understand separately. Therefore, it is natural to form PCC models with kernel functions that operate in a limited number of dimensions simultaneously.

Let $s_i(t)$ denote the observed position of individual $i \in \lbr 1, \dots, p \rbr$ at time $t \in \cT$. We model the true, unobserved position of individual $i$ at time $t$ proportional to $\tilmu_i(t)$ and add a measurement error process, $\varepsilon_i(t)$, to yield the data model
\begin{align}  
  s_i(t) &\equiv \sigma_{\mu}\tilmu_i(t) + \varepsilon_i(t). \label{eqn:s}
\end{align}
The parameter $\sigma_\mu$ scales the random process $\tilbmu(\cdot)$ to account for the overall spatial domain of the locations. We model the measurement error $\varepsilon_i(t)$ as i.i.d. mean-zero Gaussian random variables with variance $\sigma_s^2$, although other measurement error processes could also be used (e.g., \cite{Brost2015}; \citealt{Buderman2016}).

\subsection{Process prior for the dynamic social network}\label{sec:network_prior}

The general model for dependent movement is highly parameterized. To evaluate the covariance $\Delta \tau \tilbH \tilbH'$ for a grid of $m$ times, one must estimate ${p \choose 2}m$ parameters associated with the social network that drives the second stage of the PCC. As we describe in Section \ref{sec:implementation}, we fit the proposed model to data within a Markov chain Monte Carlo (MCMC) paradigm, which requires repeated inversion of a large, dense covariance matrix. Without further structure for $w_{ij}(\cdot)$, estimating the underlying network would be computationally infeasible. We increase computational efficiency by leveraging reasonable assumptions about the underlying network to constrain the space of possible latent social networks, thereby reducing the effective number of parameters in the model. We impose constraints on the network's marginal complexity at each fixed time as well as the smoothness of its temporal evolution. 

Our approach makes use of recently developed methods for modeling dynamic stochastic networks. To reduce the dimensionality of the latent network at a fixed time, we follow the approaches of \citet{Hoff2002}, \citet{Hoff2008}, and \citet{Durante2014}, who modeled social connections using a latent space. Latent space models for stochastic networks proceed by defining network connections through functions of the locations of particles in a latent $d$-dimensional ``social'' space. As an example, consider the following latent space model for the edges in a weighted, undirected, dynamic social network. Suppose the latent variables $\bmu_i^{(w)}(t)$, $\bmu_j^{(w)}(t)$, and $\bmu_k^{(w)}(t)$ are positions in $\mathbb{R}^2$, and edge weights are defined by $w_{ij}(t) = e^{-d_{ij}(t)}$, where $d_{ij}(t)$ is the Euclidean distance between $\bmu^{(w)}_i(t)$ and $\bmu^{(w)}_j(t)$. Defining edge weights based on the distance between particles in a latent space induces positive dependence among the connections $w_{ij}(t)$, because the latent distances are constrained by the characteristics of the latent space and the measure of distance. That is, if $d_{ij}(t)$ and $d_{jk}(t)$ are small, resulting in edge weights $w_{ij}(t)$ and $w_{jk}(t)$ that are close to 1, the triangle inequality implies $d_{ik}(t)$ must also be small, and $w_{ik}(t)$ will also be close to 1. Therefore, latent space approaches for modeling social networks can be understood as a way to induce a tendency to complete triangles in a network. This phenomenon is often observed in human social networks (e.g., friends of friends tend to themselves be friends; \citealt{Hansell1984}), and is also reasonable for many applications to animal social networks. A latent space approach to constraining the underlying social network has the computational advantage of reducing the number of parameters to estimate by a factor of $p$, because we only need to estimate $p$ latent paths instead of ${p \choose 2}$ pairwise relationships.

We define the latent social connection between individuals $i$ and $j$ at time $\tau_{soc}$ using the latent positions and an appropriately chosen functional $g$ by
\begin{align}  
  w_{ij}(\tau_{soc}) &\equiv g\lp \bmu_i^{(w)}(\tau_{soc}), \bmu_j^{(w)}(\tau_{soc}) \rp.
\end{align}
In general, the type of network desired for the application (e.g., binary, weighted) may inform the particular choice for $g$. As discussed in Section \ref{sec:dependent_movement}, we construct the kernel $h_{ij}^{(soc)}$ under the assumption that the weights in the social network have support $\lb 0, 1 \rb$. The compact support of $w_{ij}(\cdot)$ motivates our choice of $g$, defined by $g(\mathbf{x}, \mathbf{y}) \equiv e^{-\| \mathbf{x} - \mathbf{y} \|_2^2}$. The functional $g$ maps two vectors in the latent space to the unit interval $\lb 0, 1 \rb$ and follows the methods used for latent space network modeling by \citet{Hoff2008} who showed that this construction induces positive dependence in the edges $w_{ij}(\cdot)$.
\begin{align} 
  \bmu^{(w)}_i(\tau_w) &\equiv \sigma_w \int_\cT h^{(w)}(\tau_w, \tau)
  d\bB^{(w)}_{i}(\tau), \\
  h^{(w)}(\tau_w, \tau) &\propto e^{\frac{-(\tau - \tau_w)^2}{\phi_w^2}},
\end{align}
and $d\bB^{(w)}_{i}(\cdot)$ are instances of two-dimensional Brownian motion, now used to define the positions in the latent social space. The parameter $\sigma_w$ controls the dispersion of the latent paths $\bmu_i^{(w)}$, and is therefore related to the overall density of the social network, with smaller $\sigma_w$ corresponding to higher connectivity. The parameter $\phi^2_w$ controls the tortuosity of the trajectories through the latent social space, and is related to the temporal stability of the network over time, with larger values of $\phi_w^2$ corresponding to a more stable network. As described in Section \ref{sec:reduced_rank}, we approximate the continuous stochastic processes $\bmu_i^{(w)}(\cdot)$ using a set of independent normal random variables anchored at a finite number of knots yielding
\begin{align}  
  \bmu_i^{(w)} \sim \N \lp \bzero, \sigma_w^2 \Delta t \tilbH'_w \tilbH_w \rp.
\end{align}

\section{Model implementation}\label{sec:implementation}

We obtain realizations from the posterior distribution of the model parameters using a Markov chain Monte Carlo (MCMC) algorithm. In this section, we briefly discuss the most relevant features of model implementation. A more detailed description, including our choice of priors and hyperparameters, is available in Appendicies~\ref{app:simulation} and~\ref{app:killer_whales}.

We specify independent normal priors for the initial-location parameters and a conjugate hyperprior on the population-level variance so that $\mu_{0i} \sim \N(0, \sigma_0^2)$ and $\sigma_0^2 \sim \IG(a_0, b_0)$. Additionally, we integrate the true location process $\tilbmu$ out of the likelihood, which allows us to avoid sampling the continuous movement process directly. This is a common technique used in spatial statistics (e.g., \citealt{Gelfand2003}; \citealt{Finley2013}) to improve mixing for the remaining parameters in the model. The underlying movement process $\tilbmu$ can be recovered \textit{post hoc} using composition sampling \citep[e.g.,][]{Finley2013}. The resulting integrated model formulation is
\begin{align}  \bs_i(t) | \bep_w, \phi_{inl}, \sigma_s^2, \sigma_\mu^2, \sigma_0^2
  \sim \N \lp \bzero, \bSigma \rp \label{eqn:marginal_model} \\
  \bSigma \equiv \sigma_s^2\bI + \sigma_\mu^2 \Delta \tau \tilbH'\tilbH,
\end{align}
where the parameters $\phi_{inl}$, $\sigma_0^2$, and $\bep_w$ enter the density through the definition of $\tilbH$ in \eqref{eqn:bm} - \eqref{eqn:smooth}. To update each latent-space path, $\bmu_i^{(w)}$, we employ a Metropolis-Hastings algorithm.

\section{Simulation}\label{sec:simulation_study}

As the dependence among individuals in a population weakens (i.e., $w_{ij}(t) \rightarrow 0$ for all $t$, $i \neq j$), our proposed model simplifies to one that considers each path independently, but shares information about all non-social parameters across individuals. The PCC model constructed from a fixed, empty social network presents a natural baseline for comparison with our full model for dependent movement. Estimating the latent social network that gives rise to the dependence among paths represents the majority of the computational effort to fit our proposed model, and grows rapidly with the number of individuals under study. Thus, there is a natural incentive to model movement under the assumption of path-independence unless it can be shown to be deficient. We compare the performances of the full model for dependent movement (IP-DEP) with the special case of the model under the assumption of inter-path independence (IP-IND). In many cases, we find that accounting for the dependence among individuals results in significantly improved reconstruction of the true underlying paths. 

Telemetry devices are subject to a wide range of environmental conditions that frequently result in large intervals of time during which no observed locations are recorded. Estimates of the true locations, $\tilbmu$, during long gaps between observations are often accompanied by large amounts of uncertainty that present a challenge for researchers studying animal behavior. In simulation, we show that the presence of moderate to strong dependence among observed paths provides an opportunity for improvement in path reconstruction when we take into account the joint distribution of all observed individuals. To illustrate the potential gains, we consider a simple setting of two individuals in which the connection status, $w_{12}$, is constant over the period of observation. We assume regular, uninterrupted observations for individual 1, but a gap in time, $\cT_g$, exists in the observed sequence of telemetry locations for individual 2.

We evaluate the quality of a given model for path reconstruction in terms of both accuracy and precision. To assess the accuracy of a path reconstruction, we define a loss function termed the ``squared path error'' ($\SPE$) that quantifies the agreement between a specific path reconstruction $\hatbmu$ and the true path $\tilbmu_{true}$ as
\begin{align}  
  \SPE(\hat{\bmu}; \cT_g, \tilbmu_{true}) = 
  \frac{1}{|\cT_g|} \int_{\cT_g} \| \tilbmu_{true}(t_g) - \hat{\bmu}(t_g)\|_2^2 dt_g. \label{eqn:spb}
\end{align}
We define the reconstruction $\hatbmu$ to be the posterior mean for $\tilbmu$ so that \eqref{eqn:spb} represents a measure of how accurately the center of the posterior distribution matches the true underlying path. To assess the precision associated with a model, we compute the radii for circular 95\% credible regions surrounding each point in the path $\hatbmu(t_g)$ and average across the entire temporal gap in observations, $\cT_g$, to yield an overall summary of precision we term the average circular 95\% credible region radius ($\ACCRR_{0.95}$).

\begin{figure}[htbp]
  \centering
  \includegraphics[width = \textwidth]{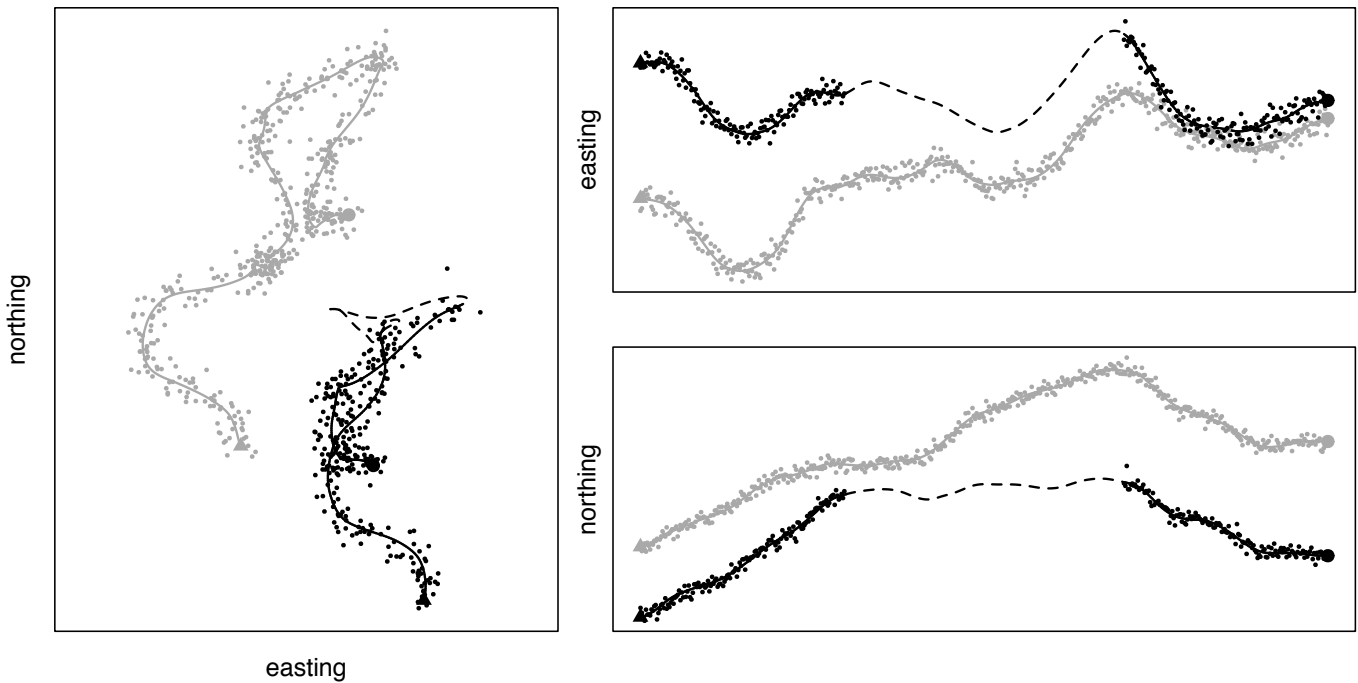}
  \caption{Example of a realization of two simulated true paths, $\tilde{\bmu}_1$ and $\tilde{\bmu}_2$ (gray and black lines, respectively), and observed locations, $\bs_1$ and $\bs_2$ (points) used in the simulation study. The dashed line represents the portion of the path taken by individual 2 where a gap in observation occurs. The figure corresponds to the case of moderate strength of tie ($w_{12} = 0.5$), large gap ($|\mathcal{T}_g|/|\mathcal{T} = 0.4|$), and low tortuosity (see also Figure~\ref{fig:pe_ratios}).}
  \label{fig:sim_eg}
\end{figure}

To provide a general sense of when accounting for path dependence results in the greatest gains in path reconstruction, we vary the strength of dependence, $w_{12}$, the proportion of the study interval made up by the observation gap, $|\cT_g|/|\cT|$, and the tortuosity of the paths, and use $\SPE$ in tandem with $\ACCRR_{0.95}$ to compare the performance of IP-DEP and IP-IND. We vary tortuosity through the range parameter $\phi_{inl}$ that appears in the inertial smoother, with smaller values corresponding to more tortuous paths on average. For each combination of $w_{12}$, $\cT_g$, and tortuosity, we used the five step procedure:
\renewcommand{\labelenumi}{\arabic{enumi}.}
\begin{enumerate}
\item Simulate a realization of $\tilbmu_{true}$ and $\bs$ from our proposed model (see Figure~\ref{fig:sim_eg}).
\item Fit both the dependent and independent-paths models to $\bs$.
\item Sample 1000 paths from the posterior distribution of $\bmu$ using composition sampling.
\item Compute the $\SPE$ defined in \eqref{eqn:spb} for the posterior mean of $\bmu_2$.
\item Use the same 1000 draws from the posterior to compute the $\ACCRR_{0.95}$ for $\bmu_2$.
\end{enumerate}
We repeat the procedure 20 times for each combination of parameters to obtain an estimate of the variability among realizations of the simulated paths. Values for the parameters used to simulate each path as well as all prior distributions and hyper parameters are provided in Table~\ref{tab:simulation}, and an example simulation is shown in Figure~\ref{fig:sim_eg}.

\begin{figure}[htbp]
  \centering
  \includegraphics[width = \textwidth]{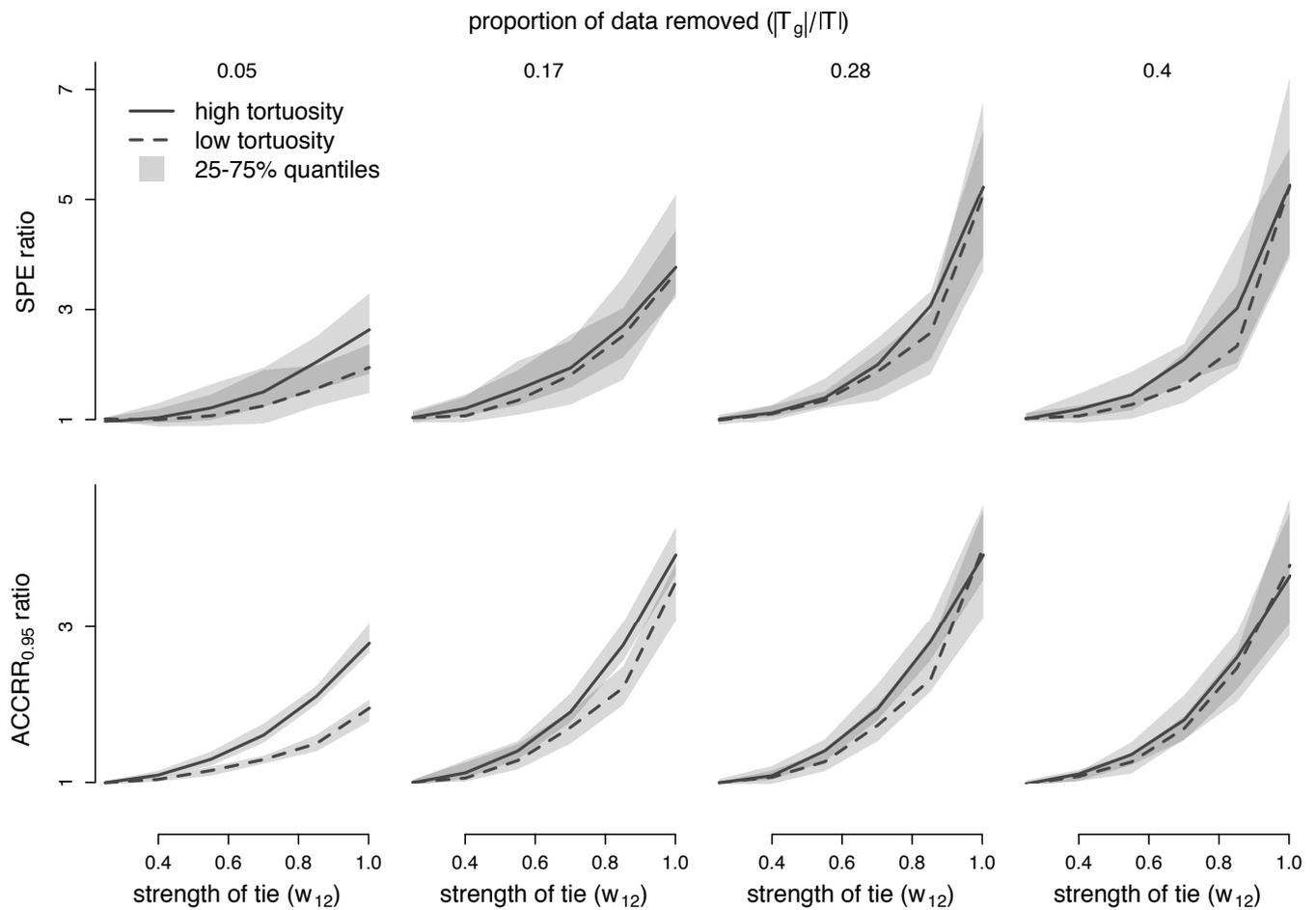}
  \caption{Path errors. The top row of plots shows the ratios of $\SPE$ for the independent (IP-IND) to full (IP-DEP) models (thus, larger ratios correspond to the full model outperforming the alternative). The bottom row shows the analogous ratios of $\ACCRR_{0.95}$. Each column represents a fixed value for the proportion of the study interval made up by the gap in observations ($|\cT_g|/|\cT|$). Within each plot, the strength of the social connection increases along the x-axis. The lines represent the median value of all ratios across the 20 simulations, and the associated polygons represent the 25\% and 75\% quantile boundaries. Finally, the solid lines correspond to simulations with high tortuosity in the true paths, while dashed lines correspond to simulations with low tortuosity.}
  \label{fig:pe_ratios}
\end{figure}

To facilitate direct comparison between the two models under consideration, we examined the ratios of $\SPE$ and $\ACCRR_{0.95}$ for IP-DEP and IP-IND, defining ratios such that the relevant value for IP-DEP appears in the denominator and values greater than 1 show support for IP-DEP. Figure~\ref{fig:pe_ratios} displays the ratios of $\SPE$ (top row) and $\ACCRR_{0.95}$ (bottom row) under all combinations of values for the parameters $w_{12}$, $|\cT_g|/|\cT|$, and $\phi_{inl}$ (tortuosity). Individual plots show the median ratio across all simulations for high (solid) and low (dashed) tortuosities, with an associated polygon delineating the 25\% and 75\% quantiles across simulations. Columns organize the plots by gap size, and the strength of the social connection $w_{12}$ increases along the x-axis within each individual plot. 

Several general observations can be made about the circumstances under which fitting the full model for dependent movement offers the greatest improvements in path reconstruction. First, both precision and accuracy improve near monotonically with increasing $w_{12}$. Second, the greatest gains come when the gap size is moderate to large (columns 3 and 4). When the gap is brief (column 1), the difference between the reconstructed paths for each model is modest. Finally, the gains in performance for the full model are greater for the case of high tortuosity. In our simulation study, more tortuous paths are characterized by a shorter range of dependence in time; thus, increasing tortuosity is similar to increasing the size of the observation gap.

\section{Killer whales}\label{sec:killer_whales}
\begin{figure}[htbp]
  \centering
  \begin{tikzpicture}
    \node[anchor=south west,inner sep=0] (image) at (0, 0) {
      \includegraphics[trim={4in 4.57in 3in 4.64in}, clip, width=\textwidth]
      {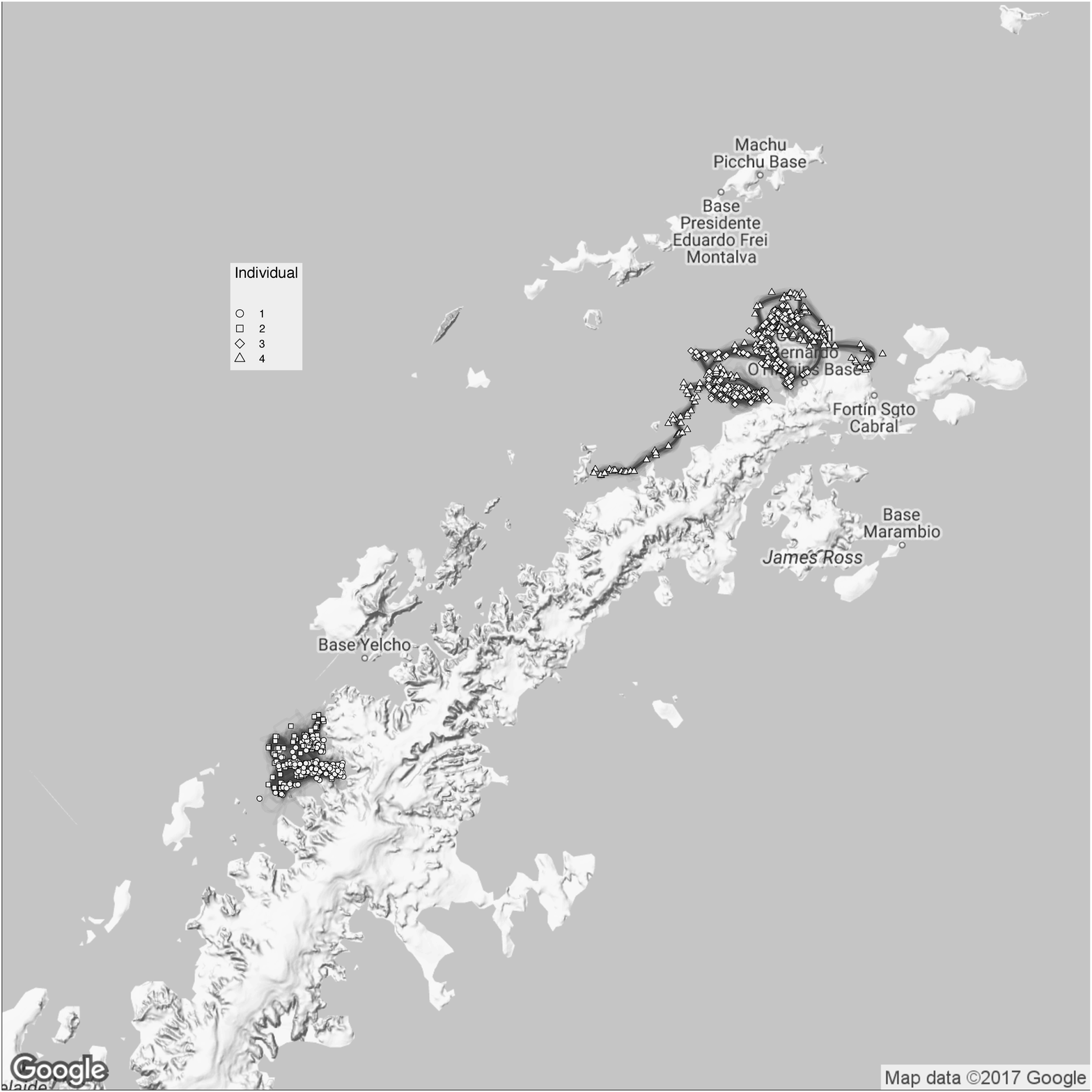}};
    \begin{scope}[x={(image.south east)},y={(image.north west)}]
      \node[anchor=south west,inner sep=0] (image) at (0.543, 0.03) {
        \includegraphics[width=0.43\textwidth]{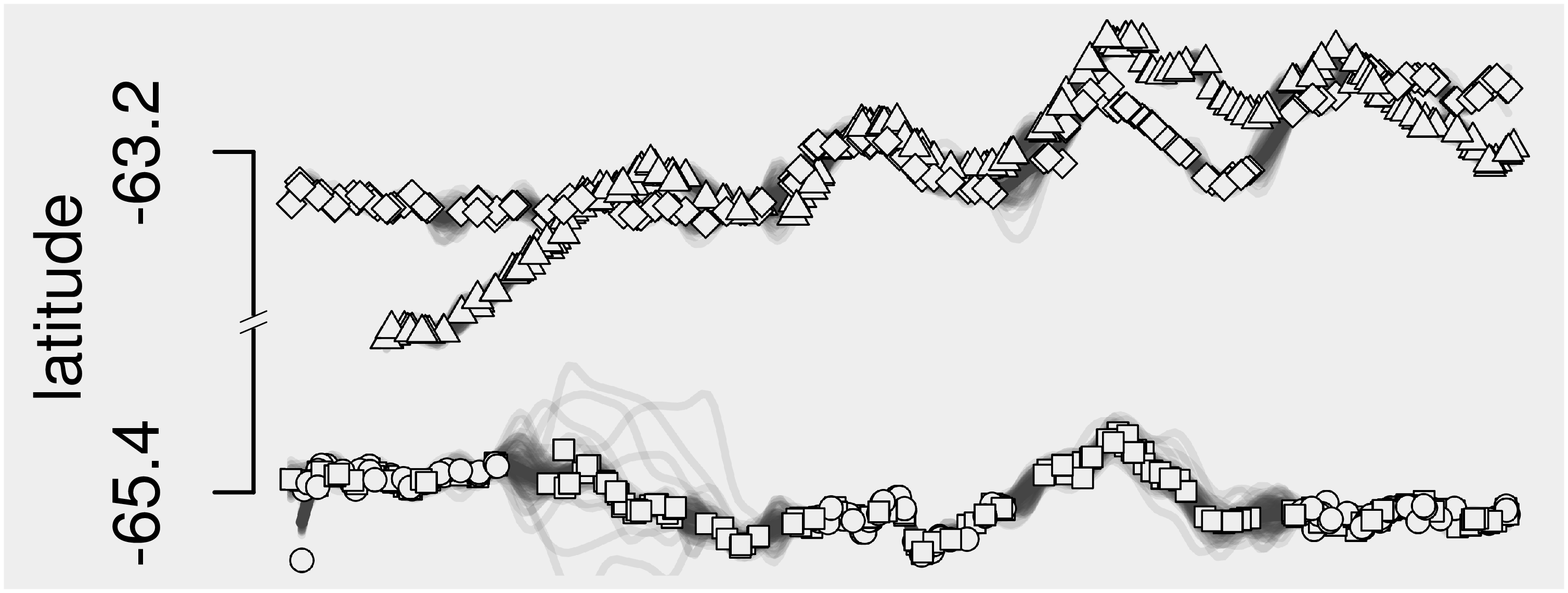}};
    \end{scope}
  \end{tikzpicture}
  \caption{Observed telemetry data and joint posterior distribution of the true paths of all individuals ($\tilbmu$). The solid lines represent the posterior mean, and the semi-transparent lines are draws from the posterior distribution to illustrate uncertainty in the path reconstruction. The points are the observed locations from the Argos satellite system. The shapes correspond to the four individuals in the study, and match those used in Figure~\ref{fig:killer_whales}. The subplot shows the latitude of each individual over time. Map created with \cite{Kahle2013}. Map data \textcopyright 2017 Google.}
  \label{fig:paths_all}
\end{figure}

We analyzed telemetry data for four killer whales near the Antarctic Peninsula (see Figure~\ref{fig:paths_all}) over the course of five days in February 2014. Geographic positions were measured using Argos telemetry tags (for a complete description of the tags and study area see \citealt{Andrews2008}; \citealt{Durban2012}). Although multiple types of killer whales have been described in this area, all four tags were deployed on individuals from the same population of the most common Type B2 killer whales \citep{Durban2016}. 

It is immediately apparent from the data that individuals 1 and 2 (Figure~\ref{fig:paths_all}, bottom left) show potential evidence of a close connection. In addition, there is some ambiguity about the relationship between individuals 3 and 4 (Figure~\ref{fig:paths_all} top right), because they occupy approximately the same spatial region during the study period. In contrast, there is little reason to suspect dependence between the two pairs of individuals. We analyzed the movement of all four individuals jointly, which allowed for pooling of information about measurement error across the entire group, and provided an opportunity for basic model validation. By fitting a model for the joint movement of all four individuals with a fully flexible latent network structure, we were able to check for the presence of potentially spurious network connections, because existing knowledge about the system suggests that the underlying network should exclude connections between the two subgroups. 

\begin{figure}[htbp]
  \centering
  \includegraphics[width = 0.67\textwidth]{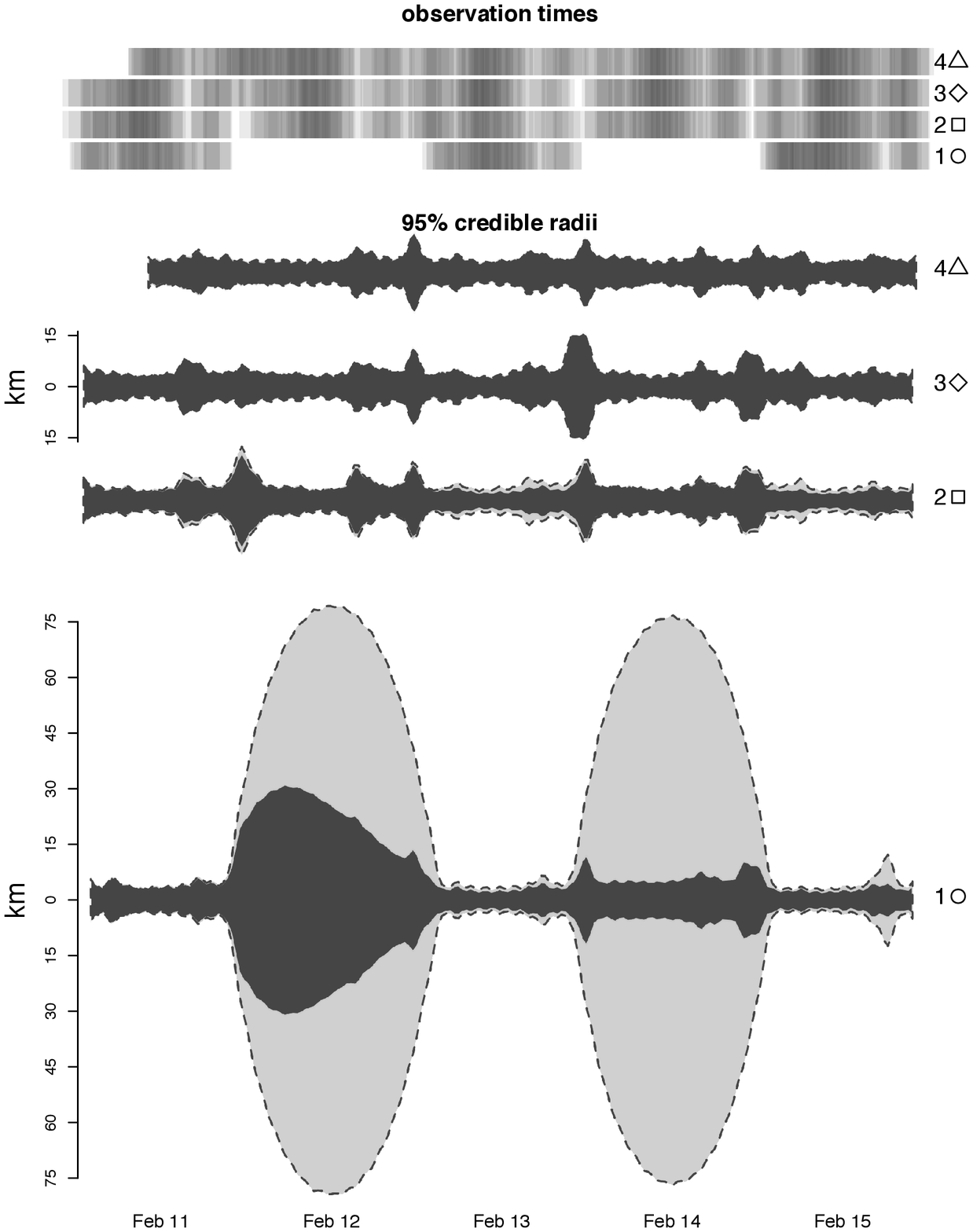}
  \caption{\textit{TOP}: Observation times for the three killer whales. Darker regions correspond to denser observation times, or equivalently, shorter gaps between observations. \textit{BOTTOM}: 95\% circular credible region radii at each time point for all four individuals. The dashed outer line shows the radii for the independent model, and the solid polygons show the radii for the full model.}
  \label{fig:killer_whales}
\end{figure}

The top plot of Figure~\ref{fig:killer_whales} shows the observation times for the killer whales in our study, with darker regions corresponding to a denser rate of telemetry measurements in time. Two different day-long gaps in observation occur for individual 1 (Figure~\ref{fig:killer_whales}), a consequence of the original study design used in the deployment of the telemetry tags which sought to balance the need for temporally dense observations with limitations in the battery life of the tags by collecting measurement only every other day for select individuals. The observation times for the other three individuals cover these gaps, suggesting that modeling the four paths jointly may allow for more precise and/or accurate estimates of the true path taken by individual 1 if moderate to strong dependence exists between it and any of the other three whales.

As in the simulation study (Section~\ref{sec:simulation_study}), we investigated path reconstructions generated by our proposed full model for dependent movement (IP-DEP), as well as the analogous model under an assumption of inter-path independence (IP-IND). We found that the full model generates path reconstructions in which the uncertainty about the position of individual 1 is dramatically reduced, compared to the reconstructions generated under the assumption of inter-path independence as measured by 95\% circular credible interval radii (bottom plot of Figure~\ref{fig:killer_whales}). Circular credible regions for the true position of the individual during the two large gaps in observation occuring on February 12th and 14th for the IP-IND model have large radii, sometimes exceeding 70km. In contrast, circular credible regions for the true position based on the IP-DEP model are less than a third this size, and similar in magnitude to the uncertainties for individuals for which we have dense observations. Despite the uncertainty about the latent social network, the IP-DEP model offers a substantial reduction in uncertainty about the true path taken by individual 1. 

\begin{figure}[htbp]
  \centering
  \includegraphics[width = 0.85\textwidth]{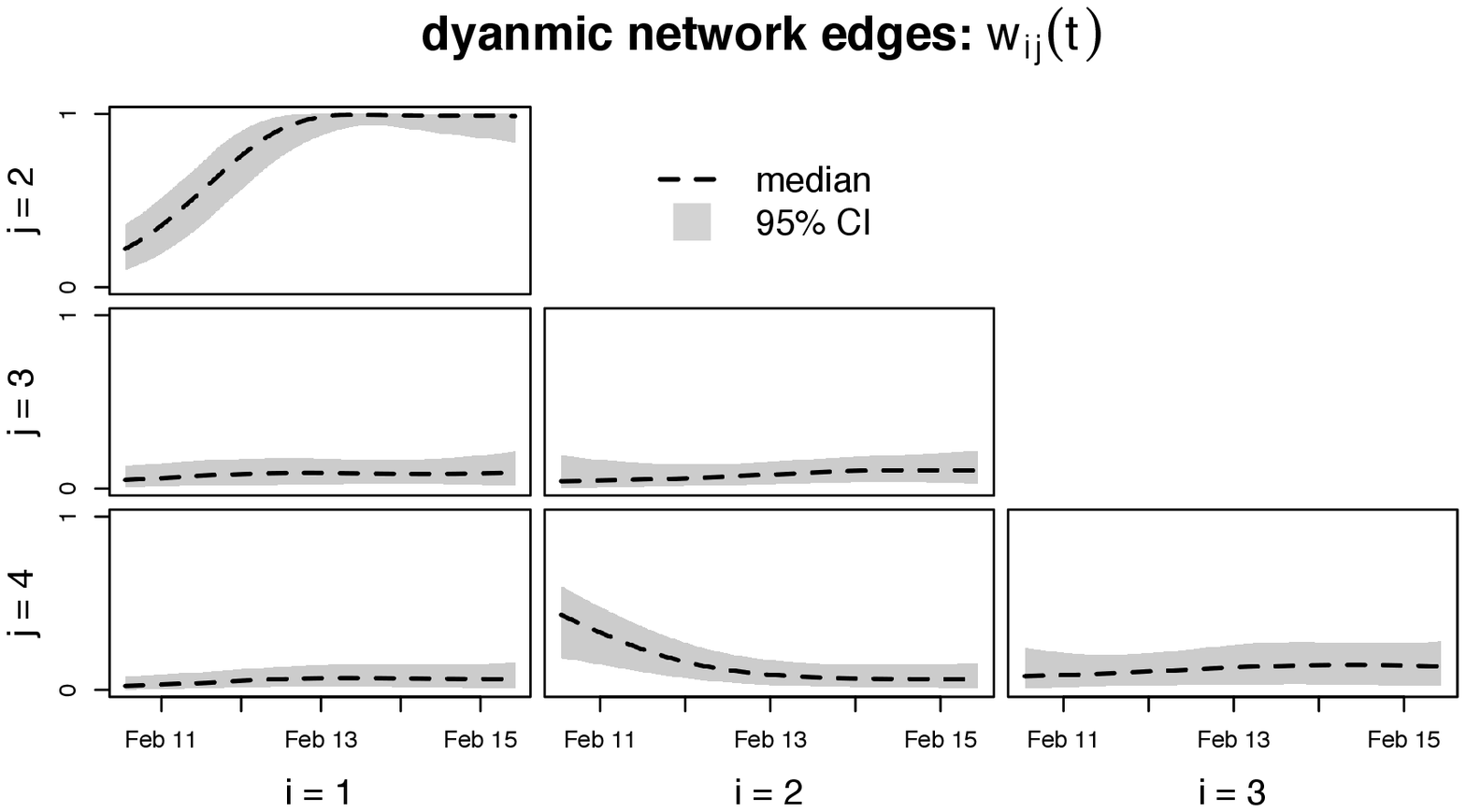}
  \caption{Posterior distribution of each edge in the dynamic, weighted social network that defines the social smoother used in stage two of the proposed PCC model (IP-DEP). Each plot shows the time-evolution for the edge, $w_{ij}(t)$, relating a specific pair of killer whales indexed by row ($i$) and column ($j$) for all times $t$.}
  \label{fig:network_posterior}
\end{figure}

The posterior distributions for the network relationships are shown in Figure~\ref{fig:network_posterior}. We can see strong evidence for a significant relationship between individuals 1 and 2, however there is weak evidence of a meaningful social connection between individuals 3 and 4. Similarly, there is no evidence of connections existing between any other pairs of individuals. Credible intervals for other model parameters, as well as the specified prior distributions and hyperparameter values are provided in Table~\ref{tab:kw_CI}. 

\section{Discussion}\label{sec:discussion}

Appropriate models for dependent, multivariate data are application-dependent. Gaussian processes offer a flexible, parsimonious tool for analyzing complex data, such as those that arise from measuring the movement of animals, if one can specify a valid form for the covariance. The need for covariances that are mathematically sound, realistic, and interpretable has motivated decades of research with a vast array of applications because satisfying all three of these goals simulataneously is challenging. Our proposed PCC framework represents a novel perspective in the construction of covariance functions. Through an application involving the movement of killer whales, we demonstrated that the PCC framework can be used to create sophisticated covariance functions that account for several important mechanisms, without resorting to unrealistic assumptions such as separability.

The joint movement of interacting individuals can be viewed as a multivariate temporal process. Methods for interpolating multivariate spatial and spatio-temporal processes, also called ``cokriging,'' have been studied for several decades (e.g., \citealt{Myers1982}; \citealt{Cressie1991}). The primary challenge has been to develop models for multivariate processes that accurately capture the dynamics within each process, as well as across processes, while ensuring the resulting cross-covariance structure is valid (symmetric and non-negative definite). While not explicitly framed as a PCC, \cite{VerHoef1998} proposed a model for cokriging that shares important connections with our proposed model for the joint movement of interacting animals. \cite{VerHoef1998} also approached the problem of specifying a valid, mechanistically motivated cross-covariance structure for multivariate processes through the use of kernel convolutions, relaxing unrealistic assumptions about the covariance structure. Similar to the way dependence among individuals arises through the application of a social smoothing kernel (Section \ref{sec:social_smoothing}), \cite{VerHoef1998} induced dependence in a multivariate process by smoothing across variables. The cross-covariance described by \cite{VerHoef1998} also accommodates spatio-temporal lags in the dependence among variables. This generalization may also be a useful feature in future models for animal movement, where a temporal lag may allow researchers to capture the effect of animals following one another. 

The particular model we constructed for animal movement cannot be used to understand all forms of dependence that may exist among interacting individuals. Rather, it has been tailored to the case in which interactions among animals manifest themselves as movement along proximally close trajectories. Particles may also exhibit other forms of dependence, such as a tendency to repel in the case of strong territorial behavior, that would not be well-described by the same covariance function. The PCC framework allows researchers to define and order kernels as necessary to appropriately model the mechanisms under study. Under a different ordering, the same three kernels we employed result in another useful covariance function for the study of particle movement. We briefly discuss this alternative to highlight the flexibility offered by the PCC framework and provide some intuition for the role certain kernel functions play in the characteristics of the resulting random process.

As mentioned in Section \ref{sec:PCC}, choosing the ordering of the kernels in a PCC represents a meaningful modeling decision when kernel convolutions do not commute. In our proposed model for dependent movement, the second and third stages of smoothing commute, while the first and second do not. We can interpret the commutativity between the social and inertial smoothers mechanistically, by noting that it does not matter whether the effect of the latent social connections operates on the raw, Brownian paths, or the temporally smoothed paths. In contrast, if we socially smoothed white noise and then used the kernel associated with Brownian motion, a very different form of dependence results. This alternative ordering ($h^{(soc)} \rightarrow h^{(bm)} \rightarrow h^{(inl)}$) presents another plausible mechanism through which dependence might arise in paths taken by multiple particles.

When applied to the velocity of a particle, smoothing with the step function kernel $h^{(bm)}$ returns the associated position process (after taking into consideration the appropriate initial location of the particle). Therefore, Brownian motion can be thought of as a random position process in which the velocity of the particle during each infinitesimal span of time is a realization from a Gaussian white noise process. In our proposed PCC model for dependent movement, the social smoother has the effect of ``shrinking'' the positions of strongly-connected particles toward each other. If we instead employed the social smoother before the step function kernel, the result would be to ``shrink'' the velocities of the particles together, rather than their positions. Thus, we would be inducing a tendency for particles to move in similar directions, though not necessarily a tendency to be in similar locations. Such an effect would be visible in the particles as movement in parallel, perhaps with a considerable distance between connected particles.

A PCC approach to constructing covariance functions for GPs allows for broad flexibility in model development, offering researchers a highly customizable framework that can be used in a wide variety of applications. We demonstrated the value of this approach with an application to animal movement, however PCCs can be used to model a broad range of random processes. Rather than relying on parametric families of covariance functions, PCCs encourage the use of interpretable, intuitive, and problem-specific convolution kernels, allowing for direct incorporation of scientific knowledge.

\section*{acknowledgements}
Killer whale tagging was conducted under permit \#14097 from the National Marine Fisheries Service and Antarctic Conservation Act permit \#2009-013. Shipboard tagging operations were supported by Lindblad Expeditions and the National Geographic Society, and by an NSF rapid grant to Ari Friedlaender. Robert Pitman helped with tag deployments and identification of killer whale types in the field.

This research was supported by NSF DMS-1614392. Any use of trade, firm, or product names is for descriptive purposes only and does not imply endorsement by the U.S. Government. The findings and conclusions of the NOAA authors in the paper are their own and do not necessarily represent the views of the National Marine Fisheries Service, NOAA.

\clearpage

\bibliography{PCC_environmetrics}

\appendix 


\section{Model specification\label{app:model}}

\subsection*{Data model}
Each orthogonal spatial direction (longitude and latitude) is modeled independently using the same specification.
\begin{align*}
  \bs &\sim \N \lp \bzero,\; \sigma_s^2 \lp 
    \bI + \sigma_{\mu/s}^2 \Delta \tau \tilbH' \bSigma_{dB} \tilbH \rp \rp \\
  \tilbH &= \bH^{(inl)}(\phi_{inl}) 
    \bH^{(soc)}(d\mathbf{B}_w, \phi_w, \sigma_w^2) 
    \bH^{(bm)} \\
  \bSigma_{dB}(t, \tau, i, j) &= \bone{t = \tau} \lp 
    \sigma_0^2 \bone{t = 0} + \bone{t > 0} \rp \bone{i = j} \\
  \bH^{(bm)}(t, \tau, i, j) &= \bone{\tau < t} \bone{i = j} \\
  \bH^{(soc)}(t, \tau, i, j) &= \bone{\tau = t} \frac{w_{ij}(t)}{|w_{i\cdot}(t)|} \\
  \bH^{(inl)}(t, \tau_{soc}, i, j) &\equiv \frac{|t - \tau_{soc}|}{\phi_{inl}}
  K_1\lp |t - \tau_{soc}| / \phi_{inl} \rp \bone{i = j} 
\end{align*}

\subsection*{Process model (integrated out for model fitting)}
\begin{align*}
  \btau & \equiv (\tau_1 = 0, \tau_2, \dots, \tau_{m-1}, \tau_m = 1)' \\ 
  \Delta \tau_i &\equiv \tau_{i} - \tau_{i - 1}, \;\; 1 < i \leq m \\
  dB(\tau_i) &\sim 
  \begin{cases} 
    \N \lp 0, \sigma_0^2 \rp; \qquad & i = 1\\
    \N \lp 0, \Delta \tau_i \rp; & i > 1
  \end{cases} \\
  d\bB &= \lp dB(\tau_1), \dots, dB(\tau_m) \rp' \\
  \tilbmu &= \bH^{(inl)}\bH^{(soc)}\bH^{(bm)}d\bB = \tilbH d\bB
\end{align*}

\subsubsection*{Prior model}
\begin{align*}
  \phi_{inl} &\sim \text{Gamma} \lp \alpha_s, \beta_s \rp 
    \quad &\sigma_0^2 &\sim \IG \lp a_0, b_0 \rp \\
  \quad \sigma_{\mu/s} &\sim \IG(a_{\mu/s}, b_{\mu/s})  
    &\sigma_s^2 &\sim \IG \lp a_s, b_s \rp
\end{align*}
\textbf{Application only:}
\begin{align*}
  \sigma_w^2 &\sim \IG(a_w, b_w)
\end{align*}

\subsubsection*{Fixed parameters}
\textbf{Simulation study:}
\begin{align*} 
  \phi_w \text{ and } \sigma_w^2 
\end{align*}
\textbf{Application:}
\begin{align*} 
  \phi_w 
\end{align*}

\section{Model fitting details} \label{app:implementation}

\subsection{Simulation Study} \label{app:simulation}

We simulated approximately continuous true paths using a grid of 500 equally spaced time points on the unit interval. We simulated 100 observation times drawn uniformly from the same unit interval. For estimation, we fixed the hyperparameters associated with the network to $\phi_w = 4/15$ and $\sigma_w^2 = 10$, and used the priors given in Table~\ref{tab:simulation}. For each fit, we obtained 4,000 iterations and discard the first 3,000 as burnin. The entire simulation study was parallelized across 6 3GHz cores and required approximately one week of computation time. 

\begin{table}[h!]
  \caption{True values and prior distibutions used in simulation study.}
  \label{tab:simulation}
  \centering
  \begin{tabular}{r|c|c}
    \hline
    parameter & true & prior density \\
    \hline
    $\phi_{inl}$ (``low tortuosity")   &  0.04     & $\text{Gamma}(2, 100)$ \\
    $\phi_{inl}$ (``high tortuosity")  &  0.04/3   & $\text{Gamma}(2, 100)$ \\
    $\sigma_0^2$      &  1      & $\IG(10^{-3}, 10^{-3})$ \\
    $\sigma_{\mu/s}^2$ & 800     & $\IG(10^{-3}, 10^{-3})$ \\
    $\sigma_{\mu}^2$   & 10      &                        \\
    $\sigma_s^2$      &  0.0125 & $\IG(10^{-3}, 10^{-3})$ \\
    \hline
  \end{tabular}
\end{table}

\subsection{Killer whales \label{app:killer_whales}}

We acquired 100,000 iterations on a single computing node, and discared the first 50,000 as burnin to yield a sample size of 50,000. We employ diffuse priors for all parameters. We set the hyperparameters associated with the network to $\phi_w = 0.3$ and $\sigma_w^2 = 10$. Model fitting was performed using a processor speed of 3 GHz and required approximately 100 hours of computing time.

\begin{table}[h!]
  \caption{Posterior credible intervals and prior distributions for the IP-DEP model in killer whale application.}
  \label{tab:kw_CI}
  \centering
  \begin{tabular}{r|rc|rc|c}
    \multicolumn{1}{c}{} & \multicolumn{2}{c}{posterior (IP-DEP)} & \multicolumn{2}{c}{posterior (IP-IND)} & prior \\
    \hline
    parameter        & median   & (2.5\%, 97.5\%)    & median  & (2.5\%, 97.5\%)    & density \\
    \hline                                                                          
    $\phi_{inl}$      & 0.00848   & (0.00756, 0.00954)   & 0.00850   & (0.00762, 0.00951)   & $\text{Gamma}(2, 100)$ \\
    $\sigma_0^2$      & 705     & (292, 2040)       & 1440     & (627, 4510)        & $\IG(1, 10)$            \\
    $\sigma_{\mu/s}^2$ & 3710     & (2600, 5490)      & 1700      & (1380, 2400)      & $\IG(10^{-3}, 10^{-3})$  \\
    $\sigma_{\mu}^2$   & 4.94    & (3.52, 7.21)       & 2.39      & (1.88, 3.14)      &  NA  \\
    $\sigma_s^2$      & 0.00133  & (0.00122, 0.00145) & 0.00133  & (0.00123, 0.00145) & $\IG(10^{-3}, 10^{-3})$   \\
    $\sigma_w^2$      & 0.491  & (0.370, 0.660) & NA  & NA & $\IG(52, 10)$   \\
    \hline
  \end{tabular}
\end{table}

\end{document}